\documentclass[twocolumn,pra,superscriptaddress]{revtex4-2}
\usepackage[utf8]{inputenc}
\usepackage{float}
\usepackage{placeins}
\usepackage{qcircuit}
\usepackage{braket}
\usepackage{amsmath}
\usepackage{amsthm}
\usepackage[normalem]{ulem}
\usepackage{todonotes}
\usepackage{xcolor}
\usepackage{multirow}
\usepackage{braket}
\usepackage{bm}
\usepackage{bbold}
\usepackage{todonotes}
\usepackage{wrapfig}
\usepackage{appendix}
\usepackage{placeins}

\newcommand{\ketbra}[2]{\mbox{$|#1\rangle\langle #2|$}}

\def\ketbra#1#2{{\vert#1\rangle\!\langle#2\vert}}

\def\braket#1#2{{\langle#1\vert#2\rangle}}

\newtheorem{theorem}{Theorem}

\newtheorem{definition}{Definition}

\usepackage{mathtools} 
\graphicspath{{figures/}}

\begin{document}

\title{Role of overparametrization in quantum approximate optimization}


\author{Daniil Rabinovich}\affiliation{Skolkovo Institute of Science and
Technology, Moscow, Russian Federation}
\affiliation{Moscow Institute of Physics and Technology,  Moscow, Russian Federation}

\author{Andrey Kardashin}
\email{Present address: Donostia International Physics Center, San Sebastián-Donostia, Spain.}\affiliation{Skolkovo Institute of Science and
Technology, Moscow, Russian Federation}

\author{Soumik Adhikary}

\affiliation{Centre for Quantum Technology, National University of Singapore, Singapore}


\begin{abstract}

Variational quantum algorithms have emerged as a cornerstone of contemporary quantum algorithms research. While they have demonstrated considerable promise in solving problems of practical interest, efficiently determining the minimal quantum resources necessary to obtain such a solution remains an open question. In this work, inspired by concepts from classical machine learning, we investigate the impact of overparametrization on the performance of variational algorithms. Our study focuses on the quantum approximate optimization algorithm (QAOA)---a prominent variational quantum algorithm designed to solve combinatorial optimization problems. We investigate if circuit overparametrization is necessary and sufficient to solve such problems in QAOA, considering two representative problems --- MAX-CUT and MAX-2-SAT. For MAX-CUT on 2-regular graphs we observe that overparametrization is both sufficient and necessary. To establish this, we analytically show that the optimal circuit depth for such problems scale as $n/2$ for even $n$. For a more general case of random graphs, the overparametrization was observed to be sufficient, yet necessary only for a statistically dominant fraction of instances. In sharp contrast, for MAX-2-SAT, underparametrized circuits suffice to solve most instances. This result highlights the potential of QAOA in the underparametrized regime, supporting its utility for current noisy devices.

 
\end{abstract}

 \maketitle

 \section{Introduction}
Noisy intermediate scale quantum (NISQ) devices are inherently limited by short coherence times, gate errors and other hardware imperfections \cite{Preskill2018}. As a result, these quantum processors can only execute short quantum circuits within a fixed error tolerance \cite{weidenfeller2022scaling,ratcliffe2018scaling,hegde2022toward, mills2022two}. This naturally limits the scale of problems that can be effectively implemented. Variational quantum computing is a novel approach designed to operate within these practical limitations by leveraging both classical and quantum computational resources \cite{cerezo2021variational, Kishore_review,mcclean2016theory_VQE}. Variational quantum algorithms (VQAs) utilize a short-depth parametrized quantum circuit (called an ansatz) which can be executed within the coherence time and does not require error correction \cite{rabinovich2024robustness, uvarov2024mitigating}. In a process reminiscent of machine learning, VQAs find a solution to a given problem by iteratively optimizing the ansatz parameters to minimize a cost function; the optimization step is executed on a classical co-processor. The specific form of the cost function is dictated by the problem. For instance, in Variational quantum eigensolvers (VQEs), which aim to prepare the ground state of a problem Hamiltonian, the cost is defined as the expectation value of the problem Hamiltonian with respect to the ansatz state.

While variational quantum algorithms (VQAs) have shown immense promise in addressing problems of practical significance \cite{farhi2014quantum, peruzzo2014variational, schuld2019quantum, bravo2023variational, adhikary2020supervised, adhikary2021entanglement}, estimating how the resources scale with respect to the ease of finding a solution remains an open challenge.
In this regard, overparametrized quantum circuits, which use a number of parameters greater than a given threshold, become promising by allowing for establishing certain convergence guarantees. 
Indeed, it has been shown analytically that for certain types of overparametrized ansatze \cite{you2022convergence} VQE converges to a small error exponentially with the number of gradient descent iterations. Moreover, in this regime, despite exponential suppression in the circuit parameter updates, the gradient descent dynamics still converges efficiently to the solution \cite{liu2024laziness, anschuetz2024unified}, avoiding the problem of barren plateaus \cite{larocca2025barren}. 
Overparametrization has also been shown to improve the success probability and convergence rate \cite{larocca2023theory} of VQE, to help in avoiding spin-glass-like cost landscapes \cite{bukov2018reinforcement, rabitz2004quantum} and to assist in mitigating the effect of noise \cite{fontana2021evaluating}. 
Importantly, overparametrization was shown to improve the probability of finding solutions to certain problems with a problem dependent ansatz \cite{larocca2023theory, you2022convergence}.

Despite showing promise in improving VQA performance, the overparametrization regime often requires circuit depth that exceed the capabilities of NISQ devices \cite{larocca2022diagnosing, larocca2023theory}. Moreover, although in certain settings some studies suggest otherwise \cite{Rosa2021BP, liu2024laziness}, such large circuit depths may hinder trainability by inducing barren plateaus. Motivated by this, we study if overparametrization is (i) sufficient and (ii) necessary for attaining solutions with a high probability for a given problem and a variational ansatz. 

It is worth noting here, that the notion of overparametrization varies across the literature; under certain definitions existing analytical results do imply sufficiency \cite{you2022convergence, liu2024laziness,  anschuetz2024unified} although the aspect of necessity remains underexplored. In this work however we follow a different definition of overparametrization \cite{haug2021capacity, larocca2023theory} using the concept of effective quantum dimension (EQD). We investigate the relationship between the circuit overparametrization depth and the optimal depth --- depths required to achieve overparametrization (saturate EQD) and to find a solution to a given problem, respectively.

We address this question with a focus on a special class of VQAs --- Quantum Approximate Optimization Algorithm (QAOA) \cite{Zhou2020, morales2020universality, lloyd2018quantum, akshay2021parameter, akshay2022circuit, Campos2021, rabinovich2022progress, dalzell2020many, Farhi2019a, Farhi2016}. It is an algorithm designed to solve combinatorial optimization problems on near-term quantum processors, utilizing a problem dependent ansatz. With long enough circuits, QAOA is capable of approximating the adiabatic evolution, thus guaranteeing success \cite{farhi2001quantum, kadowaki1998quantum,boixo2014evidence}. However no such guarantee is known for intermediate depth QAOA sequences. Thus, establishing a relationship between the optimal QAOA depth and the overparametrization depth, may offer valuable insights into performance guarantees for QAOA in the intermediate depth regime. 

We benchmark the performance of a QAOA sequence with respect to MAX-CUT problem on regular and random unweighted graphs as well as with random instances of the MAX-2-SAT problem. For the MAX-CUT problem on 2-regular graphs (a.k.a.~rings of disagrees), we analytically establish that optimal QAOA depth scales as $\lfloor \frac{n}{2} \rfloor$ for even number of qubits $n$. For odd $n$ the same scaling is confirmed numerically. Importantly, we observe that overparametrization depth exactly coincides with the optimal circuit depth for up to 20 qubits. We numerically demonstrate that this agreement approximately holds for the MAX-CUT problem on unweighted regular graphs as well as unweighted random graphs: the solutions are likely to be found exactly at (or in a close vicinity of) the overparametrization depth. The case of MAX-2-SAT problem, however,  demonstrates a different behavior: we find that the solutions to MAX-2-SAT instances could be obtained at depths much smaller than the overparametrized depth, with a high probability. 
These results demonstrate that overparametrization is sufficient to obtain solutions with a high probability. However, in sharp contrast to previous observations (e.g. \cite{larocca2023theory, you2022convergence}), it is not a necessary condition across all problems considered; while it is statistically necessary for MAX-CUT, it is not so in the case for MAX-2-SAT, thereby highlighting the potential of QAOA in the underparametrized regime.

\section{Preliminaries}
\label{sec:prelim}

\subsection{Quantum Approximate Optimization Algorithm}

The quantum approximate optimization algorithm (QAOA) is a variational quantum algorithm, originally designed to approximately solve combinatorial optimization problems \cite{niu2019optimizing,farhi2014quantum,lloyd2018quantum,morales2020universality,Zhou2020,wang2020x,Brady2021,Farhi2016,Akshay2020,Farhi2019a,Wauters2020,Claes2021}. Instances of these problems are given as pseudo-Boolean functions $\mathcal{C}: \{0,1\}^{\times n} \rightarrow \mathbb{R}$. The objective of the algorithm is to approximate a bit string that minimizes $\mathcal{C}$. To accomplish this, $\mathcal{C}$ is first encoded as a problem Hamiltonian
 \begin{equation}
      H = \sum_{{\bf x} \in \{0,1\}^{\times n}} \mathcal{C}({\bf x}) \ketbra{{\bf x}}{{\bf x}}.
 \end{equation}
The ground state of $H$ encodes the solution to the problem; in other words QAOA searches for a solution $\ket{g}$ such that $\braket{g \vert H}{g} = \min H = E_g$. 
The algorithm utilizes a $p$-depth ansatz
\begin{equation}
    \ket{\psi_p(\bm\gamma,\bm\beta)} =  \prod\limits_{k=1}^p e^{-i \beta_k H_{x}} e^{-i \gamma_k H}\ket{+}^{\otimes{n}},
    \label{ansatz}
\end{equation} 
with real parameters $\gamma_k\in[0,2\pi)$, $\beta_k\in[0,\pi)$. Here $H_x = \sum_{j=1}^{n} X_{j}$ is the standard one-body mixer Hamiltonian with Pauli  matrix $X_j$ applied to the $j$-th qubit, and $\ket{+} = \frac{1}{\sqrt{2}}(\ket{0}+\ket{1})$. The cost function is given by the expectation of the problem Hamiltonian with respect to the ansatz state. The algorithm minimizes this cost function to output:
\begin{eqnarray}
    \label{eq:approx_eng}
    &E^*_p (H)=\min_{\bm\gamma,\bm\beta} \bra{\psi_p(\bm\gamma,\bm\beta)}H \ket{\psi_p(\bm\gamma,\bm\beta)},\\
    \nonumber \\
    \label{eq:approx_sol}
    &{\bm \gamma}^*, {\bm \beta}^* \in \arg \min_{\bm\gamma,\bm\beta} \bra{\psi_p(\bm\gamma,\bm\beta)}H \ket{\psi_p(\bm\gamma,\bm\beta)}.
\end{eqnarray}

Here $E_p^* (H)$ is the estimated ground state energy of $H$ and $\ket{\psi_p({\bm\gamma}^*,{\bm\beta}^*)}$ is the estimated ground state. 
The quality of the estimation is determined by how well  $E^*_p(H)$ approximates the true ground energy $E_g$. The difference $E^*_p(H) - E_g\ge0$ is known to be a monotonically decreasing function of $p$; the problem is considered solved when this difference falls below a set threshold.
\begin{definition}[Optimal depth]
\label{optimal_depth}
    The optimal QAOA depth for a problem Hamiltonian $H$ is defined as the smallest depth $p^*$, such that $\forall p \geq p^*$, $E^*_{p}(H) - E_g \leq \epsilon $ for a given tolerance  $\epsilon \geq 0$. 
\end{definition}
As per definition~\ref{optimal_depth}, the optimal QAOA depth depends on the required tolerance $\epsilon$. While $\epsilon$ can, in principle, be chosen arbitrarily, in practice it is often set to be a fraction of the spectral gap; setting $\epsilon = \alpha \Delta$, where $\alpha \in [0,1]$ and $\Delta$ is the spectral gap, the stability lemma \cite{biamonte2021universal} guarantees $\vert \braket{\psi_p (\boldsymbol{\gamma}^*, \boldsymbol{\beta}^*)}{g} \vert^2 \geq 1-\alpha$.
An alternative approach leverages the observation that the energy error can typically undergo an abrupt drop to the machine precision, allowing one to set $\epsilon$. For the purpose of this work we set the tolerance $\epsilon=10^{-8}$, unless mentioned otherwise. 

Determining the optimal QAOA depth for an arbitrary problem, however, remains a challenging task. A typical approach involves optimizing the algorithm at a fixed initial depth and then iteratively increasing the depth until convergence is achieved. However, this method is computationally expensive.
In general, there is no efficient way to predict the optimal depth except for a restricted class of Hamiltonians. For instance, in the case of an $n$ qubit MAX-CUT problem on the ring of disagrees (2-regular graph), numerical evidence suggests that the optimal depth follows $p^* = \lfloor \frac{n}{2} \rfloor$ though a formal analytical confirmation has been lacking.


Determining the optimal depth is further complicated by the sensitivity to the initial values of $\bm \beta$ and $\bm \gamma$. Convergence to the desired accuracy heavily depends on these initial parameters. The ease of finding a solution can be quantified by the success probability, defined as a fraction of successful optimization runs (see Appendix \ref{appen:numerical_data_appendix} for details), where each run starts from different initial choices of $\bm \beta$ and $\bm \gamma$.



\subsection{Overparametrization in variational circuits}

The ability of a  variational quantum circuit to minimize a given problem Hamiltonian is intrinsically linked to its expressive power. Intuitively, this corresponds to the size of the variational state space, $\cup_{\boldsymbol{\theta}} \{\ket{\psi(\boldsymbol\theta)}\}$, generated by the corresponding ansatz circuit $U(\boldsymbol{\theta})$, applied to the initial quantum state; a larger variational state space increases the likelihood that the ground state of an arbitrary problem Hamiltonian is contained within it. The size of the variational state space, however, is not readily quantifiable. 

Alternatively, consider the distance between two variational states induced by an infinitesimal  perturbation of the circuit parameters, as defined by the Fubuni-Study metric: 
\begin{equation}
\label{eq:distance}
    D\big(\ket{\psi(\boldsymbol{\theta} +  d\boldsymbol{\theta})}, \ket{\psi(\boldsymbol{\theta})}\big) = (d\boldsymbol{\theta})^{T} \mathcal{F}(\boldsymbol{\theta}) (d \boldsymbol{\theta}).
\end{equation}
Here, $\mathcal{F}(\boldsymbol{\theta})$ is the Quantum Fisher Information (QFI) matrix \cite{liu2020quantum, yamamoto2019natural, stokes2020quantum} with elements given by:

\begin{equation}
\label{eq:QFI}
    \mathcal{F}_{jk} (\boldsymbol{\theta}) = 4 \, \mathrm{Re}\big( \braket{\partial_j \psi}{\partial_k \psi} - \braket{\partial_j \psi}{\psi} \braket{\psi}{\partial_k \psi}\big),
\end{equation} 
where $\ket{\partial_j \psi} \equiv \frac{\partial \ket{\psi(\boldsymbol{\theta})}}{\partial \theta_j}$. A non-zero distance \eqref{eq:distance} indicates that even infinitesimal perturbations to the circuit parameters result in distinct variational states. QFI is a useful tool which has various applications, including bounding the parameter estimation variance in quantum metrology \cite{sidhu2020geometric, kardashin2025predicting}, and executing the natural gradient descent in variational quantum algorithms \cite{wierichs2020avoiding, Federico2025}.
The rank of the QFI matrix $\mathcal{F(\bm \theta)}$ determines the number of orthogonal local directions in the parameter space along which infinitesimal perturbations of $\boldsymbol{\theta}$ would yield  a distinct variational state. 
The maximal rank of this matrix defines the effective quantum dimension \cite{haug2021capacity} of the variational circuit $U(\boldsymbol{\theta})$.

\begin{definition}[Effective quantum dimension]
    The effective quantum dimension (EQD) of a parameterized quantum circuit $U(\boldsymbol{\theta})$ is given by
    \begin{equation}
        \mathcal{Q}(U) = \max_{\boldsymbol{\theta}} \textup{rank}[  \mathcal{F}({\boldsymbol{\theta}})]
    \end{equation}
    \label{def:EQD}
\end{definition}

%
EQD serves as a measure of expressivity of variational quantum circuits. EQD increases monotonically with the number of circuit parameters (or circuit depth),  if the ansatz supports identity initialization, i.e. when the ansatz layers can be set to the identity operator. As EQD is bounded by the dimension of the Hilbert space, its growth necessarily saturates as additional parameters are introduced, although the onset of the saturation can happen earlier than the said upper bound, for specific cases. Notably, while definition \ref{def:EQD} requires optimization over variational parameters, in practice the optimization is rendered redundant and EQD can be computed at a random point in the parameter space \cite{haug2021capacity}. A variational circuit is considered overparametrized when increasing the number of parameters no longer increases the EQD \cite{larocca2023theory}. 
\begin{definition}[overparametrization depth]
    Overparametrization depth of a $p$-depth parametrized circuit $U_p(\bm\theta)$ is the smallest depth $p_c$ such that $\forall p \geq p_c$ the effective quantum dimension $ \mathcal{Q}(U_p) = \mathcal{Q}(U_{p_c}).$  
\end{definition}
Note that as per this definition a variational circuit is said to be overparametrized if its circuit depth is larger than $p_c$.

\section{overparametrization and optimal QAOA depth}
\label{sec:results}

As discussed in the previous section, EQD quantifies the expressivity of a variational quantum circuit and saturates in the overparametrized regime $p \geq p_c$. Furthermore, we have argued that a circuit’s ability to minimize a given problem Hamiltonian is intrinsically linked to its expressive power. This naturally raises a prudent question: what role does overparametrization play in the ability of a variational quantum circuit to minimize a given problem Hamiltonian?  

In this section, we investigate this question in the context of QAOA, focusing on instances of the MAX-CUT and the MAX-2-SAT problems. 
In particular, we investigate if overparametrization is a sufficient condition for a QAOA sequence to achieve optimal solutions and, conversely, whether it is necessary for attaining such solutions. These questions are particularly significant since both $p_c$ and $p^*$ are inherently dependent on the problem Hamiltonian, owing to the  problem-dependent ansatz used in QAOA.

\subsection{MAX-CUT on $2$-regular graphs}

The MAX-CUT problem instances on a 2-regular graph (a.k.a.~ring of disagrees) with $n$ nodes map to an $n$ qubit Ising Hamiltonian:
\begin{equation}
    \label{eq:ising_MC}
    H = \sum_{j=1}^n Z_j Z_{j+1},
\end{equation}
with a periodic boundary condition $Z_{n+1} = Z_1$. Although the ring of disagrees has an obvious solution, as suggested by the name, it serves as a valuable test bed, specially since it allows for theoretical analysis. 

\begin{figure}
    \centering
    \includegraphics[width = 0.4825\textwidth]{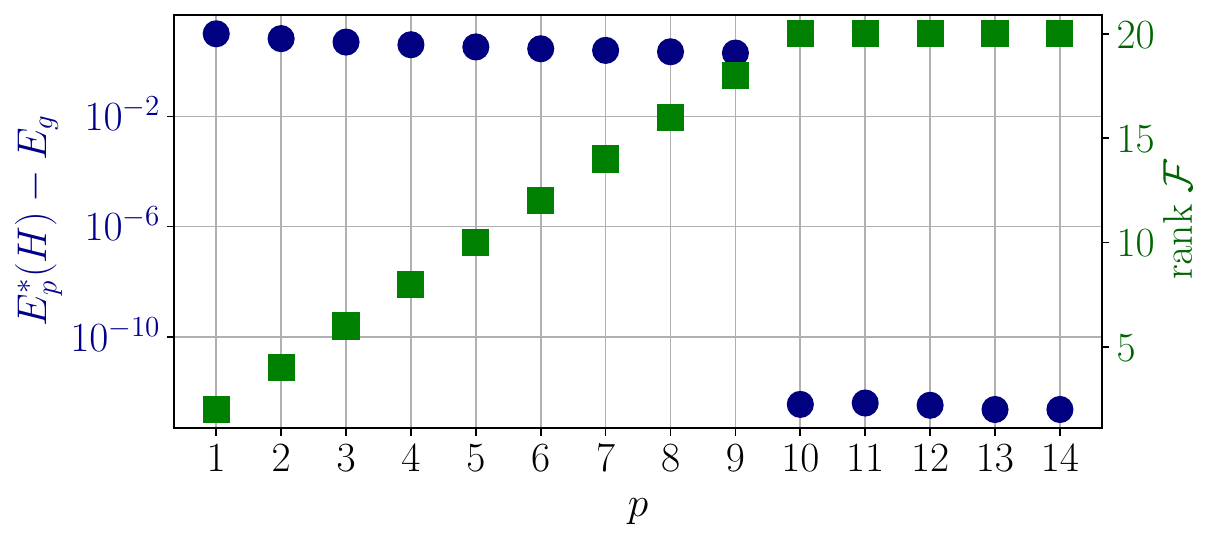}
    \caption{The energy error $E_p^*(H) - E_g$ (blue, left ordinate) and the effective quantum dimension $\mathcal{Q}(U_p)$ (green, right ordinate) with respect to QAOA circuit depth. The problem Hamiltonian $H$ is the $20$ qubit Ising Hamiltonian of type \eqref{eq:ising_MC}. 
    }
    \label{fig:20_q_rod}
\end{figure}

We begin by studying the expressivity of  the QAOA sequence for the MAX-CUT problem Hamiltonian in \eqref{eq:ising_MC}. For up to $n=20$ qubits we numerically observe that EQD grows linearly with the number of layers as $\mathcal{Q}(U_p) = 2p$ all the way up to $p_c=\lfloor \frac{n}{2} \rfloor$. Past this point, EQD stagnates and does not grow any further. A typical example is demonstrated in figure \ref{fig:20_q_rod} (green squares). 

Interestingly, previous numerical studies on the ring of disagrees have reported that the optimal depth also scales as $p^*=\lfloor \frac{n}{2} \rfloor$ \cite{fermionic}, which has been conjectured to hold for any system size $n$. Indeed, we numerically reproduce this result for up to $n=20$ qubits, reaching the tolerance $\epsilon\sim10^{-12}$. This is demonstrated in figure \ref{fig:20_q_rod} for $n=20$ (blue circles), where a sharp drop in  $E^*_{p}(H) - E_g$ is observed at $p = 10$. The observed drop in $E^*_{p}(H) - E_g$ is characteristic to the problem and in Appendix \ref{appen:drop} we argue that the drop remains finite even in the thermodynamic limit for even values of $n$. For odd values of $n$ however the drop can be shown to vanish in the limit $n \rightarrow \infty$. 
Building on our numerical observations, we provide an analytical guarantee for the optimal QAOA circuit depth. 

\begin{theorem}[Optimal QAOA depth]
    Let $H$ be the MAX-CUT problem Hamiltonian on a 2-regular graph with an even number of qubits $n$. Then, for $p = \frac{n}{2}$, $E^*_{p}(H) - E_g = 0$.
    \label{theorem}
\end{theorem}
The proof is presented in Appendix \ref{appen:proof_ising}, where we propose a set of angles $({\bm \beta}, {\bm \gamma})$ with $p = \frac{n}{2}$, such that the corresponding state satisfies $E_{p}(H) = \bra{\psi_p(\bm\gamma,\bm\beta)}H \ket{\psi_p(\bm\gamma,\bm\beta)} = E_g$, hence proving that the proposed set of angles is optimal. While this result does not rule out the possibility of finding a solution at a lower depth, it indeed  establishes an upper bound $p^* \leq \frac{n}{2}$.

Notably, we also observe a sharp transition in the success probability at $p = p^* = p_c = \frac{n}{2}$, further indicating the ease of finding solutions to the ring of disagrees problem at the overparametrization depth. This agreement between the overparametrization depth and the optimal QAOA depth is remarkable, as it indicates that not only is overparametrization sufficient to find solutions with high probability, it is also necessary.

\subsection{MAX-CUT on random graphs}
\label{app:maxcut_random}
To further substantiate the role of overparametrization, we systematically investigate the performance of QAOA across a broader class NP-hard MAX-CUT problem for graph sizes of up to $n=10$ nodes. Given an arbitrary graph $G=(V,E)$, the MAX-CUT problem can be encoded into an $n=|V|$ qubit problem Hamiltonian
\begin{equation}
    H = \sum_{(i,j) \in E} Z_i Z_j,
\end{equation}
whose ground state encodes the solution to the problem.
We consider both $k$-regular graphs---a natural generalization of the 2-regular graphs considered in the previous section--- with $k \in [3,7]$ and unweighted random graphs. 
For both cases of random and regular graphs, qualitatively similar results are obtained. Thus, the section focuses on the results obtained for the unweighted random graphs, which represent a richer class of problems.

\begin{figure}[h]
\center{\includegraphics[width=1\linewidth]{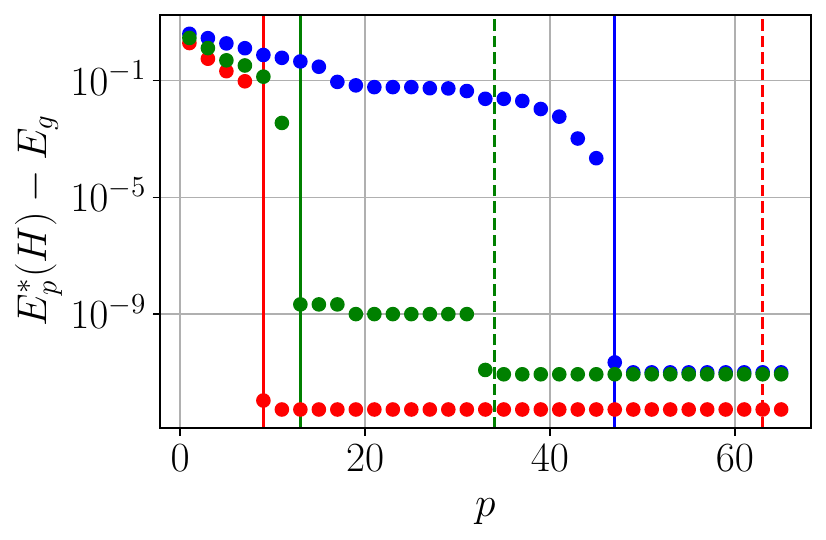}}  \\
\caption{Typical examples of energy errors for $n=7$ vertex random graphs for $q=0.6$. The best energy, found in 20 optimization runs, is depicted as a function of circuit depth.
The solid vertical lines represent the optimal depth $p^*$, while the dashed lines represent the overparametrization depth $p_c$. 
}
\label{fig:maxcut-random}
\end{figure}

\begin{figure}[!tbh]
 \begin{minipage}[b]{\linewidth}
   \centerline{\includegraphics[clip=true,width=3.2in]{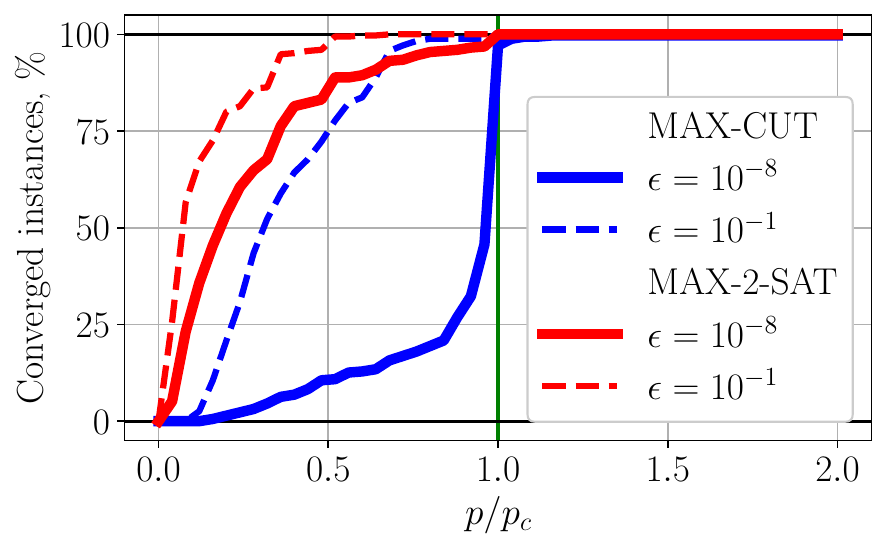}} (a)
   \end{minipage}
   \begin{minipage}[b]{\linewidth}
   \centerline{\includegraphics[clip=true,width=3.2in]{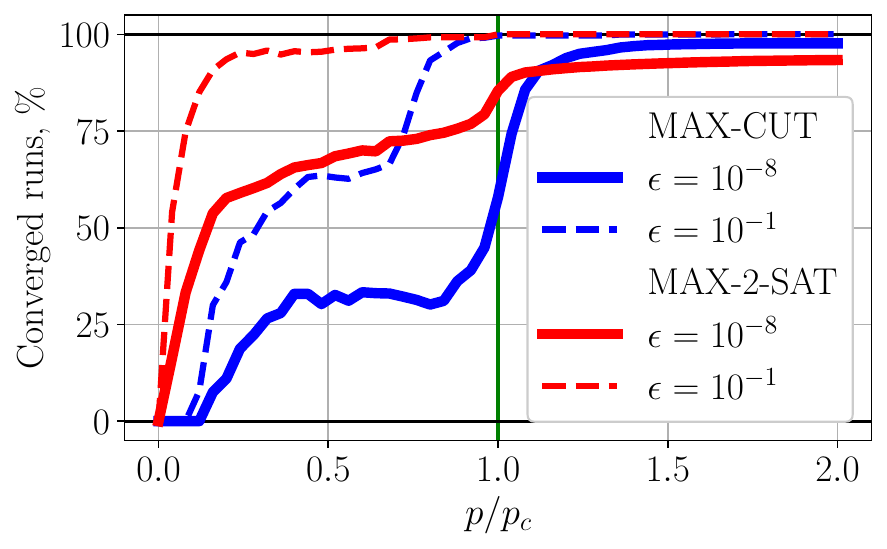}} (b)
   \end{minipage}
   \caption{Optimization results for MAX-CUT (in blue) and MAX-2-SAT (in red) on $n=7$ qubit instances. Solid and dashed lines depict the convergence for $\epsilon = 10^{-8}$ and $\epsilon=10^{-1}$ energy error thresholds, respectively. (a) Fraction of solved instances with respect to the normalized circuit depth $p/p_c$. The fraction is computed over all the considered instances (50 random instances for each edge probability $q$, and 50 random instances for each number of clauses $m$, for MAX-CUT and MAX-2-SAT, respectively). (b) Fraction of converged optimization runs for the instances considered in (a) with respect to the normalized circuit depth.}
   \label{fig:sat_and_cut}
 \label{helix}
 \end{figure}

For each graph size $n$ and each edge probability $q \in [0.3, 0.9]$ varied with a step size of $0.1$, we sample $50$ random instances. Each instance is then minimized using QAOA with the layerwise heuristic strategy (this choice is motivated in Appendix \ref{appen:numerical_data_appendix})
using $20$ optimization runs with random parameter initializations. Figure \ref{fig:maxcut-random} illustrates different types of energy behavior with respect to circuit depth. The blue curve depicts the case when the solution is found exactly at the overparametrization depth $p_c$. 
The green and red lines, on the other hand, illustrate that overparametrization is not necessary for preparing the exact ground state for certain problem instances. 
However, a systematic comparison of the optimal and overparametrization circuit depth demonstrates that most of the instances are solved exactly at or slightly before the overparametrization depth $p_c$ (similar to the blue curve in figure \ref{fig:maxcut-random}). Indeed, this is demonstrated in the solid blue curve of figure \ref{fig:sat_and_cut}(a) which indicates an abrupt increase in the fraction of solved instances, as one transitions from the underparametrized to the overparametrized regime. Note that, here the normalized depth $p/p_c$ is used as it enables a uniform comparison of QAOA performance across problem instances with varying overparametrization depths. 

The importance of overparametrization becomes clearer if success probability is taken into account. To accomplish this, for each value of $p/p_c$ we compute the average success probability over all instances that were solved by the considered normalized depth. The solid blue curve in figure ~\ref{fig:sat_and_cut}(b) demonstrates the variation of this success probability with respect to the normalized circuit depth. One can clearly observe that the success probability experiences an abrupt increase when transitioning from the underparametrized to the overparametrized regime. Combining the observations from figures \ref{fig:sat_and_cut}(a) and (b) we conclude that overparametrization is sufficient to solve a MAX-CUT instance with a high probability. Conversely, while overparametrization is not a necessary condition, a major portion of the problem instances get solved solely in the immediate vicinity of $p/p_c = 1$. Moreover, the success probability of finding the solutions experiences a significant improvement only as one transitions into the overparametrized regime.

Nevertheless, it is worth noting that the critical role of overparametrization becomes less pronounced for larger tolerance $\epsilon$ in definition \ref{optimal_depth}. In dashed blue lines in figure \ref{fig:sat_and_cut} we demonstrate how fast the instances and optimization runs converge to the error tolerance $\epsilon=0.1$, which show a dramatic improvement compared to lower tolerance. Note that the solution can now be extracted only probabilistically, as single shot measurement will extract the exact solution with at least $1-\epsilon/\Delta\ge 0.9$ probability. However, this would typically be considered a tolerable cost, as it allows to significantly reduce circuit depth---the main obstacle for quantum computing in the NISQ era.

\subsection{MAX-2-SAT}
In this section we investigate how the observed relationship between the overparametrization and optimal circuit depth generalizes to other combinatorial problems. To this end we turn our attention to a different class of NP-hard combinatorial optimization problems --- MAX-2-SAT. Given a boolean formula written in conjunctive normal form with exactly $k$ literals per clause, a MAX-k-SAT problem consists of fining a variable assignment that maximizes the number of satisfied clauses. The problem can be embedded into a Hamiltonian minimization problem for \begin{equation}
    H_{\text{SAT}}=\sum_j P(j),
    \label{eq:SAT}
\end{equation}
where $j$ indexes clauses of an instance, and $P(j)$ is the tensor product of projectors that penalizes bit string assignments that do not satisfy the $j$-th clause. When rewritten in the Pauli basis, \eqref{eq:SAT} becomes a $k$-local Ising type Hamiltonian. In this work we focused on the $k=2$ case.

In the numerical experiments, we consider MAX-2-SAT instances with $n \in [4,10]$ variables and a number of clauses $m \in [2,2n]$ with a step size of $2$, for each value of $n$. The choice of $m$ was dictated by the observation that overparametrization depth $p_c$ stagnates at the maximum of $2^n-1$ for the clause density $\alpha=m/n \gtrsim 2$, with a similar stagnation reported for the optimal circuit depth \cite{akshay2022circuit}. $50$ instances of random, uniformly generated MAX-2-SAT problem were considered for each combination of $n$ and $m$. Each instance is minimized using QAOA with the layerwise heuristic strategy using $50$ optimization runs with random parameter initializations.

In a striking contrast to the observation for MAX-CUT, most of the MAX-2-SAT instances were  solved well within the underparametrized region, $p/p_c \ll 1$. This is demonstrated in the solid red curve of figure \ref{fig:sat_and_cut}(a) which shows the fraction of solved MAX-2-SAT instances with respect to the normalized depth. This result indicates that overparametrization is not necessary to solve instances of the MAX-2-SAT problem. Importantly, as highlighted by the solid red curve in figure~\ref{fig:sat_and_cut}(b) the success probability also becomes significant in the underparametrized regime, even at $p/p_c\ll1$. This is in stark contrast to our observation for MAX-CUT, where for $\epsilon=10^{-8}$ the success probability remained low at $p/p_c \ll 1$, even though overparametrization was not deemed necessary to solve the instances. As expected, lowering the threshold $\epsilon$ for MAX-2-SAT improves the convergence even further, see dashed red lines in figure \ref{fig:sat_and_cut}.

\begin{figure}
    \centering
    \includegraphics[width = 0.495\textwidth]{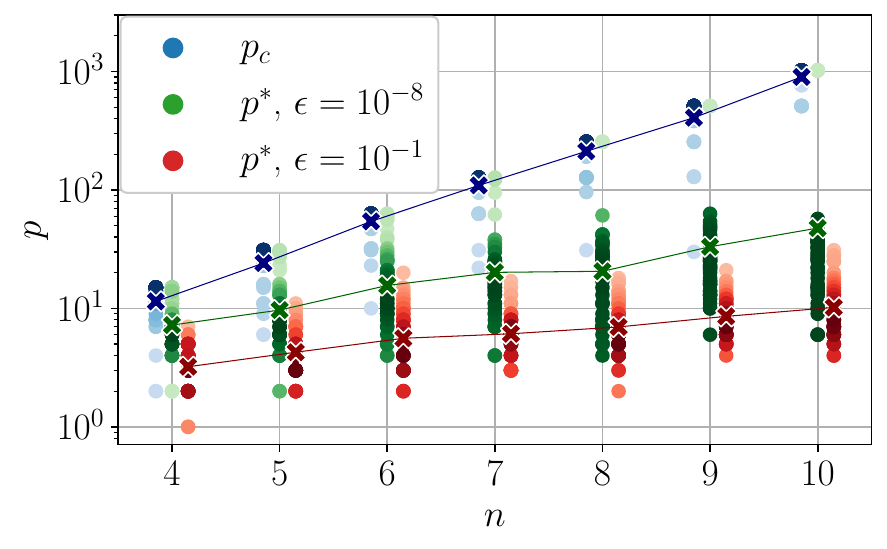}
    \caption{Comparison of the overparametrization and optimal QAOA circuit depth scaling for MAX-2-SAT problems of different size $n$ and clause density $\alpha = 2$. The statistics is accumulated over 50 random instances for each problem size $n$. The color intensity represents the density of data points with the respective depth. The crosses depict the geometric means of the corresponding depths. The data is offset horizontally for better visualization.}
    \label{fig:depth_scaling}
\end{figure}

The above observations suggest that the overparametrization depth serves as a reasonable estimate for the depths required to solve instances of the MAX-CUT problem to a negligible energy error. In contrast, for MAX-2-SAT, the correlation between the optimal and the overparametrization depth is less pronounced. Motivated by this, we investigate the scaling of both $p^*$ and $p_c$ with respect to the size of the problem instances $n$. Out of $350$ instances considered in our experiment, only $21$ were not solved by the depth $p=150$; for these instances, we set $p^* = 2^n-1$. The results can be seen in figure~\ref{fig:depth_scaling} for problem instances of density $\alpha = m/n = 2$. First, it can be observed that the overparametrization threshold $p_c$ scales with the dimension of the Hilbert space, as has been established for other notions of overparametrization \cite{liu2024laziness, you2022convergence}. Interestingly, in this case $p_c$ almost saturates the dimension, as $p_c\sim2^n$. The optimal circuit depth, despite seemingly exhibiting exponential behavior as well, demonstrates more modest growth, $p^*\sim\sqrt{2^n}$ for the energy error threshold $\epsilon=10^{-8}$. As a result, the difference between the two is clearly seen to increase exponentially with problem size $n$, thereby confirming the observation that a majority of MAX-2-SAT instances can be solved well within the underparametrized regime. This observation becomes even more pronounced for a higher energy error threshold, $\epsilon=10^{-1}$, as depicted in red in figure \ref{fig:sat_and_cut}. Interestingly, the observed scaling implies that upon problem size increase, the instances are getting solved by increasingly more underparametrized circuits, as $p^*/p_c \sim 1/\sqrt{2^n}$.


\section{Conclusion and discussion}
\label{sec:conc}

In classical machine learning, overparametrization has been extensively studied and is known to improve both approximation and generalization performance~\cite{belkin_overparam, allen2019learning}. These insights have motivated analogous investigations for VQAs, owing to their structural similarities with classical machine learning models such as neural networks. 
Existing studies tend to focus on the overparametrized quantum circuits \cite{you2022convergence, liu2024laziness, anschuetz2024unified} which allow establishing theoretical convergence guarantees, proving sufficiency under certain definitions of overparametrization for solving problems with specific ansatz types.
A more quantifiable approach for defining overparametrization was introduced by \cite{larocca2023theory} using the concept of effective quantum dimension \cite{haug2021capacity}. In this work the authors demonstrate empirical evidence for sufficiency as well as necessity of overparametrization, by showing convergence improvement upon transitioning into the overparametrized regime. In our work we adopted the same EQD-based definition of overparametrization threshold and investigate whether it is indeed both sufficient and necessary for finding exact solution with high probability.
We addressed this question in the setting of QAOA with respect to distinct classes of NP-hard combinatorial optimization problems --- MAX-CUT and MAX-2-SAT. 

For both problem classes, we found that overparametrization serves as a sufficient condition for obtaining exact solutions with a high probability. For MAX-CUT on 2-regular graphs, we further established that it also serves as a necessary condition---the overparametrization depth exactly matches the analytically determined optimal circuit depth, which we prove to scale linearly with the problem size $n$ for even $n$. This agreement approximately holds for other graph classes ($k$-regular and random graphs): most instances were solved either exactly at, or in close proximity to the overparametrization depth. The role of overparametrization is further highlighted when considering the success probability, quantified as the fraction of optimization runs that converge to the optimum. In good agreement with \cite{larocca2023theory} the success probability exhibited a pronounced increase upon transitioning from the underparametrized to the overparametrized regime. Taken together, these results indicate that, for MAX-CUT instances, the overparametrization depth offers a reasonable estimate for the optimal depth needed to achieve solutions with high probability. 

The MAX-2-SAT problem, despite having a Hamiltonian formulation reminiscent of MAX-CUT, exhibited markedly different behavior. A majority of instances were solved well within the underparametrized regime, $p/p_c \ll 1$. 
To put this in perspective, $\sim 85\%$ of the instances with $n=7$ were solved using circuits with depths as small as half the overparametrization depth. Likewise, most optimization runs converged deep in the underparametrized regime, resulting in a significantly high success probability even at $p/p_c \sim 1/\sqrt{2^n}$. This behavior stands in stark contrast to the findings of~\cite{larocca2023theory}, where the success probability remains negligible and optimization stagnates in the underparametrized regime; the probability of finding the solution experiences an abrupt increase only when the number of circuit parameters approaches the dimension of the dynamical Lie algebra (DLA), i.e.~when transitioning to the overparametrized region.

This behavior also differs markedly from related results for the Quantum Walk Optimization Algorithm (QWOA), a generalized version of QAOA, where it was shown that for all problems belonging to the class NP Optimization with Polynomially Bounded objective value (NPO-PB) but not belonging to the class BPPO (Bounded-error Probabilistic Polynomial Time Optimization) require highly overparametrized QWOA sequence to be solved \cite{QWOA_Adhikary}. One arrives at a similar observation for the case of the unstructured search problem~\cite{akshay2021parameter, grover1996fast} with QAOA; the dimension of DLA (and hence the overparametrization depth $p_c$) is upper bounded by $O(n^2)$ due to the ansatz symmetries, while the optimal circuit depth is lower bounded by $O(\sqrt{2^n})$, due to the optimality of Grover scaling \cite{grover1996fast}. 
Nevertheless, we believe the findings of the current work are largely encouraging. The fact that overparametrization is not necessary to solve MAX-2-SAT instances highlights the potential of QAOA in the underparametrized regime.

The potential of the underparametrized circuits becomes further pronounced when noting that, due to their reduced depth they are less prone to noise, making them more favorable for NISQ devices. Moreover, this indicates that certain problems can be solved with substantially fewer resources than would otherwise be needed if overparametrization was necessary. Nevertheless, the circuit can suffer from trainability limitations, such as barren plateaus, already in the underparametrized regime. Indeed, cost function gradients can become exponentially small already at the polynomial depth $p=poly(n)$ \cite{cerezo2021cost}, while the overparametrization depth typically scales with the dimension of Hilbert space, 
$p_c\propto 2^{n}$. Thus, for large $n$ it is natural to expect that gradients become suppressed even in the underparametrized regime with normalized depth $p/p_c\ll1$.
Further improvement in performance can potentially be expected upon considering  other modified versions of QAOA such as in \cite{vijendran2024expressive}, which face no restrictions induced by barren plateaus. 

\vspace{1em}
\section*{Acknowledgments}
The work of D.R. and A.K. was supported by Rosatom in the framework of the Roadmap for Quantum computing (Contract No. 868-1.3-15/15-2021 dated October 5, 2021 and Contract No. №R2163 dated December 03, 2021). The large calculations were performed on the Zhores supercomputer \cite{zacharov2019zhores}.

\section*{Data availability}
The data that support the findings of this article are openly available at \cite{qaoa-qfi-code}.


\appendix

\onecolumngrid




\section{Abrupt energy drop for the ring of disagrees}
\label{appen:drop}
The cost function $E^*_{p}(H) - E_g$ for the ring of disagrees experiences an abrupt drop at depth $p^*$, as demonstrated in figure \ref{fig:20_q_rod}. Here we argue that the cost function experiences a finite drop from $p=p^*-1$ to $p=p^*$ even in the thermodynamic limit for even $n$, similar to the case depicted in the main text.

Note that the optimal value of the energy at any depth $p<p^* = \lfloor \frac{n}{2}\rfloor$ is observed to be 
\begin{eqnarray}
    E_p^*(H) = -n\frac{p}{p+1}.
\end{eqnarray}
This scaling agrees with the results of \cite{farhi2014quantum, fermionic}, and we numerically confirm it up to $n=20$. The energy improvement between the consecutive layers in close vicinity of the optimal depth $p\lesssim p^*$ is then given by
\begin{eqnarray}
    E_{p-1}^*(H)-E_{p}^*(H) \sim \frac{4}{n},
\end{eqnarray}
which becomes small for a large problem size. For the odd number of qubits, the energy improvement at the last layer
\begin{equation}
    E^*_{p^*-1}(H) - E^*_{p^*}(H) = \frac{2}{n-1}\sim \dfrac{2}{n},
\end{equation}
is of the same order and vanishes for $n\to \infty$. Here we used that the optimal energy at $p^*$ equates $-(n-2)$. At the same time, the optimal energy for even number of qubits is $E^*_{\frac{n}{2}}(H) = -n$. Hence, the energy drop at the optimal depth is 
\begin{equation}\label{eq:cons_impr}
    E^*_{p^*-1}(H) - E^*_{p^*}(H) = 2
\end{equation}
and remains finite for any even number of qubits $n$. Thus, we conclude this behaviour to be a fundamental feature of the ring of disagrees problem, remaining even in the thermodynamic limit.

\section{Proof of Theorem~\ref{theorem}}
\label{appen:proof_ising}

In this section we prove that a QAOA circuit of depth $p=\frac{n}{2}$ finds the exact solution to the ring of disagrees problem \eqref{eq:ising_MC} of even values of $n$; $E_p^* (H) = E_g = -n$. For this, we propose a two distinct set of parameters---which differ slightly, depending on the parity of $\mathrm{mod}(n,4)$--- at which one attains the absolute minimum of $E_p(\bm \beta, \bm \gamma)$. We begin by considering $n=4m+2$ for integer $m$ and show that the exact minimum $E_p^*(H) = -n$ is achieved at the parameters
\begin{align}
    \label{eq:angles_opt_ising_1}
    &\beta_k = \gamma_k = \frac{\pi}{4}; \ \  k \in \left[1:p\right], k \neq m+1 \nonumber \\
    &\beta_{m+1} = \frac{\pi}{8}, 
    ~~~~~\gamma_{m+1} = \frac{3 \pi}{8}.
\end{align}
To prove it we shall show that
\begin{equation}
    E_p(H)\equiv
    \bra{+}^{\otimes n}V^{\dagger m} 
    U(\gamma,\beta)^\dagger V^{\dagger m} H V^m 
    U(\gamma,\beta) V^m \ket{+}^{\otimes n} = n \sin{4\gamma}\sin{4\beta},
    \label{eq:expectation}
\end{equation}
where $p = \frac{n}{2}$ and $U(\gamma,\beta)  \equiv e^{-i \beta H_x} e^{-i \gamma H}$ is a single QAOA layer. For the sake of brevity we have denoted $\beta_{m+1} = \beta$, $\gamma_{m+1} = \gamma$ and $U\Big(\frac{\pi}{4},\frac{\pi}{4}\Big) = V$ in \eqref{eq:expectation}. Thereafter, it is straightforward to see that $E_p(H)$  attains its minimal value $-n$ when $\beta=\pi/8$, $\gamma=3\pi/8$ (or vice versa). Note that the problem Hamiltonian \eqref{eq:ising_MC} is invariant under cyclic permutation operations hence ensuring $\bra{\psi_p}H\ket{\psi_p}=n\bra{\psi_p}Z_k Z_{k+1}\ket{\psi_p}$. 
In order to prove \eqref{eq:expectation} we make use of Pauli backpropagation by conjugating the original $Z_kZ_{k+1}$ term with QAOA circuit. Notably, as $\bra{+}Z\ket{+}=\bra{+}Y\ket{+}=0$, it suffices to keep track of the Pauli strings that will be composed only of identities and Pauli $X$ operators by the end of backpropagation.  

We begin by establishing how certain Pauli strings transform under the conjugation. These rules are straightforward to obtain, yet they involve lengthy algebraic manipulations. Specifically, under conjugation with $V$, a Pauli string of $l$ Pauli X operators with $Z$ and/or $Y$ at the edges transforms as

\begin{align}
& V^{\dagger} Z_k X_{k+1}\dots X_{k+l} Z_{k+l+1}V =
\begin{cases}
Z_{k-1}X_k X_{k+1}\dots X_{k+l} X_{k+l+1} Z_{k+l+2},~ 0\le l \le n-4\\
X_k X_{k+1}\dots X_{k+l} X_{k+l+1}, l = n-3\\
Y_k X_{k+1}\dots X_{k+l} Y_{k+l+1}, ~l = n-2\\
\end{cases}
\label{eq:ZXXZ} 
\\
& V^{\dagger} Y_k X_{k+1}\dots X_{k+l} Y_{k+l+1}V  =   \begin{cases}
Y_{k+1} X_{k+2}\dots X_{k+l-1} Y_{k+l}, 2\le l \le n-2\\
-X_{k+1}, l = 1\\
Z_k Z_{k+1}, l = 0
\end{cases}
\label{eq:YXXY} 
\\
&V^{\dagger} Z_k X_{k+1}\dots X_{k+l-2}X_{k+l} Y_{k+l+1}
V  =
 Z_{k-1} X_k X_{k+1} \dots X_{k+l-2}   Y_{k+l}\mathbb{1}_{k+l+1},
\label{eq:ZXXY}
\\
&V^{\dagger} Y_k X_{k+1} X_{k+2}\dots X_{k+l} Z_{k+l+1}
V  =
 \mathbb{1}_k Y_{k+1} X_{k+2}\dots X_{k+l} X_{k+l+1} X_{Z+l+2}.
\label{eq:YXXZ}
\end{align}
In this notation $O_{k+n}\equiv O_k$ for any operator $O$. In other words the rules \eqref{eq:ZXXZ}-\eqref{eq:YXXZ} mean that the string of $X$ operators grows from the $Z$-edge and shrinks from the $Y$-edge, until one of the special cases in \eqref{eq:ZXXZ} and \eqref{eq:YXXY} is met. To simplify the explanation, we refer to strings on the left hand sides of equations \eqref{eq:ZXXZ}-\eqref{eq:YXXZ} as ZX...XZ, YX...XY, ZX...XY and YX...XZ strings in the respective order. Notice that according to the aforementioned rules only the ZX...XZ and YX...XY strings with an odd number of $X$ operators can produce strings solely composed of $X$ and identity operators---the only strings that could contribute to the ultimate expectation value. Armed with these rules, we are set to prove \eqref{eq:expectation}.

Starting with the $Z_kZ_{k+1}$ and applying the rule \eqref{eq:ZXXZ} one establishes that 
\begin{equation}
    H_1\equiv V^{\dagger m} Z_k Z_{k+1} V^m = Z_{k-m} X_{k-m+1}\dots X_{k+m} Z_{k+m+1},
    \label{eq:H_1}
\end{equation}
and subsequently obtains
\begin{align}
    H_2\equiv U(\gamma,\beta)^\dagger H_1    U(\gamma,\beta) = 
    &\cos^2 2\beta \cdot e^{2i\gamma (Z_{k-m}Z_{k-m+1}+Z_{k+m}Z_{k+m+1})} \cdot Z_{k-m} X_{k-m+1}\dots X_{k+m} Z_{k+m+1}\nonumber\\+
    &\sin^2 2\beta \cdot e^{2i\gamma (Z_{k-m-1}Z_{k-m}+Z_{k+m+1}Z_{k+m+2})} \cdot Y_{k-m} X_{k-m+1}\dots X_{k+m} Y_{k+m+1}\nonumber\\+
     &\frac{1}{2}\sin 4\beta \cdot e^{2i\gamma (Z_{k-m-1}Z_{k-m}+Z_{k+m}Z_{k+m+1})} \cdot  Y_{k-m} X_{k-m+1}\dots X_{k+m} Z_{k+m+1}\nonumber\\+
     &\frac{1}{2}\sin 4\beta
     \cdot e^{2i\gamma (Z_{k-m}Z_{k-m+1}+Z_{k+m+1}Z_{k+m+2})} \cdot Z_{k-m} X_{k-m+1}\dots X_{k+m} Y_{k+m+1}.
     \label{eq:gammabeta_layer}
\end{align}
The expression \eqref{eq:gammabeta_layer} is obtained by commuting $U$ to the left of the string, next to $U^\dagger$; upon this procedure a large number of terms got self-eliminated as $U$ commutes with the bulk of the string \eqref{eq:H_1}. 
Notice that the first two lines of \eqref{eq:gammabeta_layer}, when opening the exponents, will produce only the YX...XZ and ZX...XY strings, alongside with the ZX...XZ and YX...XY strings with an even number of $X$ operators. As explained above, these strings will not contribute to the expectation value upon further conjugation with $V$. Following the same logic, when expanding the exponents in the last two lines of \eqref{eq:gammabeta_layer}, the ZX...XY and YX...XZ strings can be committed, yet the resultant ZX...XZ and YX...XY strings will have an odd number and are the only strings that have to be considered. Thus, the only contributing terms in \eqref{eq:gammabeta_layer} read
\begin{align}
    H_2^{\text{contr}}&=
     \frac{i}{4}\sin 4\gamma\sin 4\beta \cdot ( Z_{k-m-1}Z_{k-m}+Z_{k+m}Z_{k+m+1}) \cdot Y_{k-m} X_{k-m+1}\dots X_{k+m} Z_{k+m+1}\nonumber\\
     &+\frac{i}{4}\sin 4\gamma\sin 4\beta
     \cdot (Z_{k-m}Z_{k-m+1}+Z_{k+m+1}Z_{k+m+2}) \cdot Z_{k-m} X_{k-m+1}\dots X_{k+m} Y_{k+m+1}\nonumber\\
     &=\frac{1}{4}\sin 4\gamma\sin 4\beta(Z_{k-m-1}X_{k-m} X_{k-m+1}\dots X_{k+m} Z_{k+m+1}+Z_{k-m} X_{k-m+1}\dots X_{k+m} X_{k+m+1}Z_{k+m+2})-
     \nonumber\\
     &-\frac{1}{4}\sin 4\gamma\sin 4\beta(Y_{k-m} X_{k-m+1}\dots X_{k+m-1} Y_{k+m}+ Y_{k-m+1} X_{k-m+2}\dots X_{k+m} Y_{k+m+1})
     )
     \label{eq:H_2^contr}
\end{align}
Once again, notice that the first two strings have $2m+1$ $X$ operators and, once conjugated with $V^m$, will end up in the second case of \eqref{eq:ZXXZ}. Similarly, the last two terms in \eqref{eq:H_2^contr} have $2m-1$ $X$ operators, thus will end up in the second case of \eqref{eq:YXXY}. In other words, 
\begin{align}
    V^{\dagger m} H_2^{\text{contr}} V^m\nonumber = \frac{1}{4}\sin 4\gamma\sin 4\big(X_{k-2m}\dots X_{k+2m}\mathbb{1}_{k+2m+1}+X_{k-2m+1}\dots X_{k+2m+1}\mathbb{1}_{k+2m+2}+X_k+X_{k+1}\big).
\end{align}
Thus, the energy
\begin{align}
    E_p(H)&=n \bra{+}V^{\dagger m} H_2^{\text{contr}} V^m\nonumber\ket{+} = n \cdot \sin4\beta\sin4\gamma
    \label{E_P(H)_final}
\end{align}
indeed has a global minimum at $\gamma = 3\pi/8$ and $\beta =\pi/8$ (or vice-versa) with the corresponding energy being $E_p^* (H) = -n$. The obtained energy exactly coincides with the ground energy of the considered Hamiltonian, hence ensuring the exact minimum. The remaining case $n = 4m$ can be considered in a similar fashion, and one can follow the same steps as established above to show that the exact minimum $E_p^*(H) = -n$ is achieved at the parameters
\begin{align}
    &\beta_k = \frac{\pi}{4}; \ \  k \in \left[1:p\right], k \neq m \nonumber \\
     &\gamma_k = \frac{\pi}{4}; \ \  k \in \left[1:p\right], k \neq m+1 \nonumber \\
    &\beta_{m} = \frac{\pi}{8}, ~~~~\gamma_{m+1} = \frac{3 \pi}{8}.
\end{align}
Interestingly, in both cases of $n=4m$ and $n=4m+2$ the circuit is composed of multiple exponents with angles $\pi/4$, with the two distinct angles sitting exactly in the middle of the circuit---whether in the same layer or in the neighboring ones. Interestingly, similar patterns in optimal parameter were observed for a few of $4$-regular graphs, yet no relation to the ring of disagrees was found. These results can be conveniently established using certain graphical rules for track Pauli backpropagation, which we leave for future works.

\section{Optimization strategy and data post-processing}
\label{appen:numerical_data_appendix}

There may be different heuristic strategies for optimization in QAOA \cite{sud2024parameter, sack2023recursive,sack2021quantum}. Among the most prominent are the ones described in \cite{Zhou2020}.
        There, the authors propose to reparametrize the ansatz angles via the so-called Fourier encoding, or interpolate the parameters found for $p$ ansatz layers to $p+1$ layers.
        Similarly to the latter, Ref.~\cite{lee2023depth} introduces a bilinear initialization strategy for the $(p+1)$-th layer, which utilizes the parameters for the layers $p-1$ and $p$.
        More broadly, the optimal parameters can be transferred even between different problem instances within a problem class (e.g., between different graphs in the MAX-CUT problem) \cite{brandao2018fixed, wurtz2022counterdiabaticity}.

        While many optimization strategies for QAOA exist, we intend to test the performance of QAOA in relation to the overparametrization depth $p_c$, regardless of any heuristics applied. Therefore, the most appropriate strategy for our purposes would be using no heuristic strategy whatsoever. That is, ideally, we should have optimized each instance of the QAOA ansatz for each $p$ from scratch with random initializations.
        However, this would consume a lot more computational time.

        Therefore, in our work, we use a strategy similar to the one implemented in \cite{akshay2022circuit}.
        Namely, we start the optimization with $p=1$ ansatz layer.
        Once the optimization terminates, we use the optimal parameters $(\beta_1^*, \gamma_1^*)$ to initialize the optimization with $p=2$ layers at the parameters $(\beta_1^*, \beta_2^{(0)}, \gamma_1^*, \gamma_2^{(0)})$, where $\beta_2^{(0)}$ and $\gamma_2^{(0)}$ are sampled uniformly. All the parameters are then optimized simultaneously.
        This procedure is repeated until reaching a desired number of layers $p$ or attaining the target accuracy.

       To this end, the simple strategy of re-using the previous angles found for $p-1$ layers to initialize the $p$-layered QAOA serves as a compromise between saving computational resources and not introducing too much enhancement via heuristics.

\begin{figure}
         \begin{minipage}{0.475\linewidth}
           \centerline{\includegraphics[width=\linewidth]{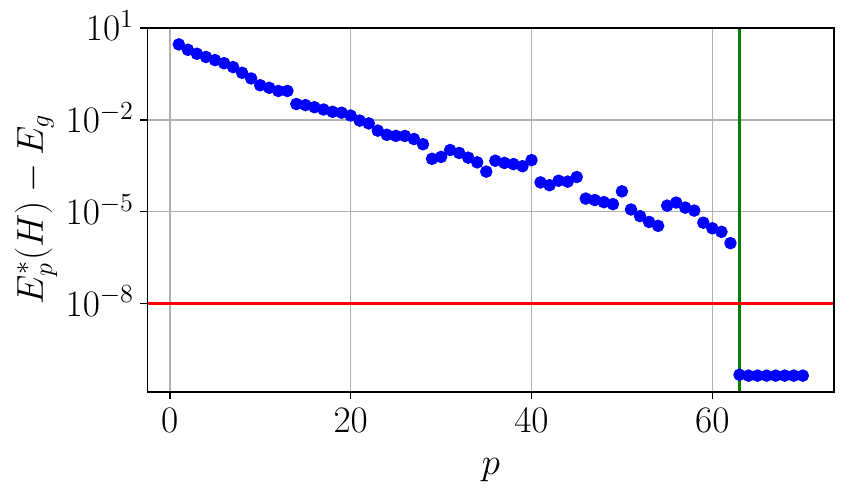}} (a)
           \end{minipage}
           \hspace{0.5cm}
           \begin{minipage}{0.475\linewidth}
           \centerline{\includegraphics[width=\linewidth]{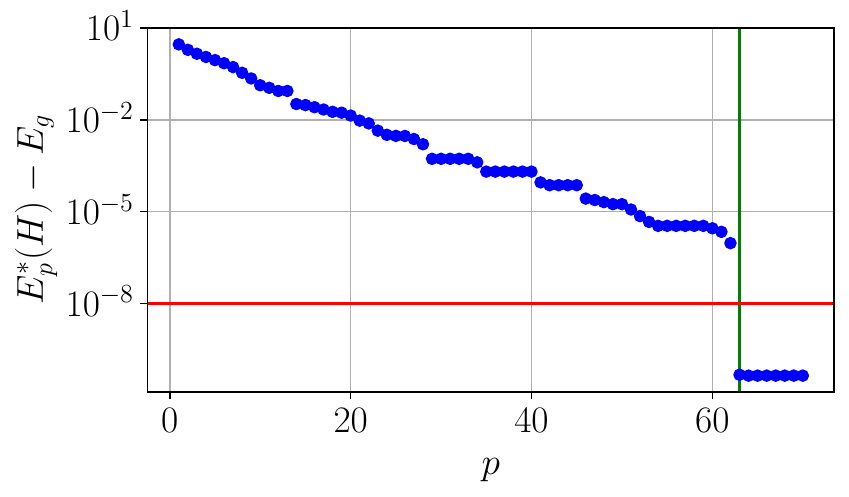}} (b)
           \end{minipage}
           \caption{Energy errors for a MAX-CUT instance on a random graph of  $n=7$ vertices with $q=0.5$ as function of depth $p$.
           (a) The smallest error found in 20 optimization runs at each $p$. 
           Due to the optimization strategy used, the error achieved at $p + 1$ layers may be large compared to $p$ layers.
           (b) When this is the case, we set $E_{p+1}^*(H) = E_{p}^*(H)$, which ensures monotonic behaviour of the error.
           In both panels, the green line indicates $p_c$ (which coincides with $p^*$ in this case), and the red line stands for the error threshold $\epsilon = 10^{-8}$.
           }
           \label{fig:errors-noncum-cum}
         \end{figure}

        Although this strategy may save computational time, much like any other initialization strategy, it has certain drawbacks. As each new layer is initialized randomly, the optimization with $p+1$ layers does not necessarily converge to lower cost function values compared to the $p$ layer optimization.
        That is, the relation $E_{p+1}^*(H) \leq E_{p}^*(H)$ may not hold for this strategy, as the optimization may stuck in a higher local minimum.
        Therefore, we post-process the data by setting $E_{p+1}^*(H) = E_{p}^*(H)$ whenever we obtained a higher energy error at $p+1$ layers compared to $p$ layers.
        This is demonstrated in figure~\ref{fig:errors-noncum-cum}, where we show the minimum energy error obtained for a MAX-CUT instance across $20$ optimization runs with and without the post-processing described above.
        Note that this post-processing is also applied in figure~\ref{fig:maxcut-random} in the main text.
Importantly, usage of more sophisticated heuristic training strategies might provide better starting points and thus improve convergence. This would only reduce the reported optimal depth $p^*$, making the underparametrized regime even more favorable. 

\FloatBarrier

        A similar post-processing is used when we count the optimization runs converged to a given error threshold $\epsilon$.
        Due to the reasons mentioned earlier, we may have fewer converged runs with $p+1$ than with $p$.
        However, we can enforce a non-decreasing behavior here as well.
        Namely, if a run converged to the threshold $\epsilon$ with some depth $p_\mathrm{conv}$, we consider it to be converged for all $p \geq p_\mathrm{conv}$ as well. With this strategy in place  the success probability---defined as the fraction of converged optimization runs in the main text---allows one to keep track of the number of runs that have converged at the considered depth or before. The same can be said about the instances: if one of the runs converged at $p_\mathrm{conv}$, then the instance is deemed to be solved for all $p \geq p_\mathrm{conv}$.
        This post-processing was implemented in figure~\ref{fig:sat_and_cut} in the main text.





\end{document}